\documentclass[acmsmall]{acmart}
\usepackage{hyperref,cleveref,subcaption}
\usepackage{wrapfig}
\AtBeginDocument{%
  \providecommand\BibTeX{{%
    Bib\TeX}}}

\setcopyright{acmcopyright}
\copyrightyear{2023}
\acmYear{2023}
\acmDOI{XXXXXXX.XXXXXXX}

\acmJournal{JACM}
\acmVolume{37}
\acmNumber{4}
\acmArticle{111}
\acmMonth{8}

\AtBeginDocument{%
  \providecommand\BibTeX{{%
    \normalfont B\kern-0.5em{\scshape i\kern-0.25em b}\kern-0.8em\TeX}}}

\newtheorem{dfn}{Definition}

\begin{document}

\title{On the Opportunities and Challenges of Offline Reinforcement Learning for Recommender Systems}

\author{Xiaocong Chen}
\email{xiaocong.chen@data61.csiro.au}
\affiliation{%
  \institution{Data61, CSIRO}
  \country{Australia}
}
\author{Siyu Wang}
\email{siyu.wang5@unsw.edu.au}
\affiliation{%
  \institution{UNSW Sydney}
  \country{Australia}
}
\author{Julian McAuley}
\email{jmcauley@eng.ucsd.edu}
\affiliation{
\institution{UCSD}
\country{USA}
}
\author{Dietmar Jannach}
\email{dietmar.jannach@aau.at}
\affiliation{
\institution{University of Klagenfurt}
\country{Austria}
}
\author{Lina Yao}
\email{lina.yao@unsw.edu.au}
\affiliation{%
  \institution{Data61, CSIRO \& UNSW Sydney}
  \country{Australia}
}

\renewcommand{\shortauthors}{Chen et al.}

\begin{abstract}
Reinforcement learning serves as a potent tool for modeling dynamic user interests within recommender systems, garnering increasing research attention of late. However, a significant drawback persists: its poor data efficiency, stemming from its interactive nature. The training of reinforcement learning-based recommender systems demands expensive online interactions to amass adequate trajectories, essential for agents to learn user preferences. This inefficiency renders reinforcement learning-based recommender systems a formidable undertaking, necessitating the exploration of potential solutions. Recent strides in offline reinforcement learning present a new perspective. Offline reinforcement learning empowers agents to glean insights from offline datasets and deploy learned policies in online settings. Given that recommender systems possess extensive offline datasets, the framework of offline reinforcement learning aligns seamlessly. Despite being a burgeoning field, works centered on recommender systems utilizing offline reinforcement learning remain limited. This survey aims to introduce and delve into offline reinforcement learning within recommender systems, offering an inclusive review of existing literature in this domain. Furthermore, we strive to underscore prevalent challenges, opportunities, and future pathways, poised to propel research in this evolving field.
\end{abstract}



\received{20 February 2007}
\received[revised]{12 March 2009}
\received[accepted]{5 June 2009}

\maketitle

\section{Introduction}
In recent years, notable advancements have materialized in the realm of recommendation techniques, transcending the scope of traditional approaches (such as collaborative filtering, content-based recommendation, and matrix factorization~\cite{lu2015recommender}). This evolution has led to the emergence of deep learning-based methods in the field of recommender systems (RS). The appeal of deep learning stems from its ability to comprehend intricate non-linear relationships between users and items, making it adept at accommodating diverse data sources like images and text.
The adoption of deep learning in RS has proven beneficial in tackling multifaceted challenges. Its strength lies in addressing intricate tasks and managing complex data structures~\cite{zhang2019deep}.

Traditional recommendation systems (RS) have limitations in capturing interest dynamics, a challenge that emphasizes the distinction between users' long-term and short-term interests~\cite{chen2020knowledge,zhang2019deep}. Specifically, while these traditional methods are adept at recognizing and modeling long-term interests based on historical data and patterns, they often fall short in accounting for the rapidly changing and more nuanced short-term interests. This gap in responsiveness to short-term shifts can lead to recommendations that are out-of-sync with a user's current preferences or situational needs. 
In contrast, deep reinforcement learning (RL) aims to train an agent with the capacity to learn from interaction trajectories provided by the environment, achieved through the integration of deep learning and RL techniques as expounded in \cite{chen2023deep}. Notably, this approach empowers the agent to proactively glean insights from real-time user feedback, thereby enabling the discernment of evolving user preferences within the dynamic context of reinforcement learning.

RL provides a structured mathematical framework for acquiring learning-based control strategies. By employing RL, we can systematically attain highly effective behavioral policies, which encapsulate action strategies. These policies are engineered to optimize predefined objectives referred to as reward functions. In essence, the reward function serves as a directive, guiding the RL algorithm towards desired actions, while the algorithm itself devises the means to enact these actions. Throughout its history, the field of RL has been a subject of intensive research. More recently, the integration of robust tools like deep neural networks into RL methodologies has yielded substantial advancements. These neural networks act as versatile approximators, empowering RL techniques to exhibit exceptional performance across a diverse array of problem domains.

Nevertheless, a pertinent challenge to the widespread implementation of RL techniques emerges. RL methods fundamentally follow an incremental learning approach, wherein they gather knowledge by iteratively engaging with their environment. Subsequent refinements are informed by previous experiences. While this iterative learning approach is effective in numerous scenarios, its practicality is not universal. Consider cases such as real-world robotics, educational software pedagogy, or healthcare interventions; these situations entail potential risks and resource expenses that cannot be disregarded.
Moreover, even within scenarios conducive to online learning, such as in the context of RS, a preference for historical data often arises. This preference is particularly pronounced in intricate domains where sound decision-making hinges upon substantial data inputs. The rationale is that leveraging previously amassed data enables informed decisions without necessitating continuous real-world experimentation.

The success of machine learning methods in solving real-world problems in the past decade is largely thanks to new ways of learning from large amounts of data. These methods get better as they're trained with more data. However, applying this approach to online Reinforcement Learning (RL) doesn't fit well. While this wasn't a big problem when RL methods were simpler and used small datasets for easy problems, adding complex neural networks to RL makes us wonder if we can use the same data-driven approach for RL goals. This would mean creating a system where RL learns from existing data without needing more data collected in real-time~\cite{levine2020offline}.

However, this idea of using existing data for RL brings its own challenges. As we discuss in this article, many common RL methods can learn from data collected differently from how the policy behaves. But these methods often struggle when trying to learn effectively from a whole set of data collected in advance, without more data being collected as the policy improves. Making things more complicated with high-dimensional neural networks can make this problem worse. A big issue with using pre-existing data for RL is that the data's distribution may not match real-world conditions~\cite{levine2020offline}. Still, the potential of a fully offline RL system is exciting. Just like how machine learning can turn data into useful tools like image recognition or speech understanding, an offline RL system, using strong function approximators, might turn data into smart decision-makers. This could mean that people with lots of data could make policies that help them make better choices for what they want to achieve~\cite{mazoure2022improving}.

RS and advertising are particularly well-suited areas for applying offline RL. This is because collecting data is straightforward and efficient, often done by recording user actions. Moreover, the existing RS literature provides sufficient datasets which can be used for training offline RL. However, these domains are also critical in terms of safety. Making a very poor decision could lead to significant financial losses. Therefore, traditional online exploration methods are not practical here. This is why offline RL methods have a history of being used in these fields.

One technique commonly employed is called off-policy evaluation. This approach is useful for running A/B tests and estimating the effectiveness of advertising and RS methods without needing to interact with the environment further.

In the case of RS, things are a bit different compared to other applications. RS policy evaluation is often set up as a contextual bandit problem. Here, "states" might represent a user's past behavior, and "actions" are the recommendations made to them. This simplification avoids the complexity of sequential decision making, which is useful. However, it can lead to inaccuracies if actions are connected over time, like in robotics or healthcare scenarios.

Using offline RL for RS has practical applications such as optimizing recommendations presented together on a page, improving entire web pages, and estimating website visits with the help of doubly robust estimation. Another use is A/B testing to fine-tune click rates for optimization. Researchers have also used offline data to learn policies. This includes efforts like improving click-through rates for newspaper articles, ranking advertisements on search pages, and tailoring ad recommendations for digital marketing.

In this survey, our main focus will be on offline RL in RS (offline RL4RS). We aim to provide a comprehensive overview of existing works, along with discussing open challenges and future directions.

\begin{figure}
    \centering
    \includegraphics[width=0.8\linewidth]{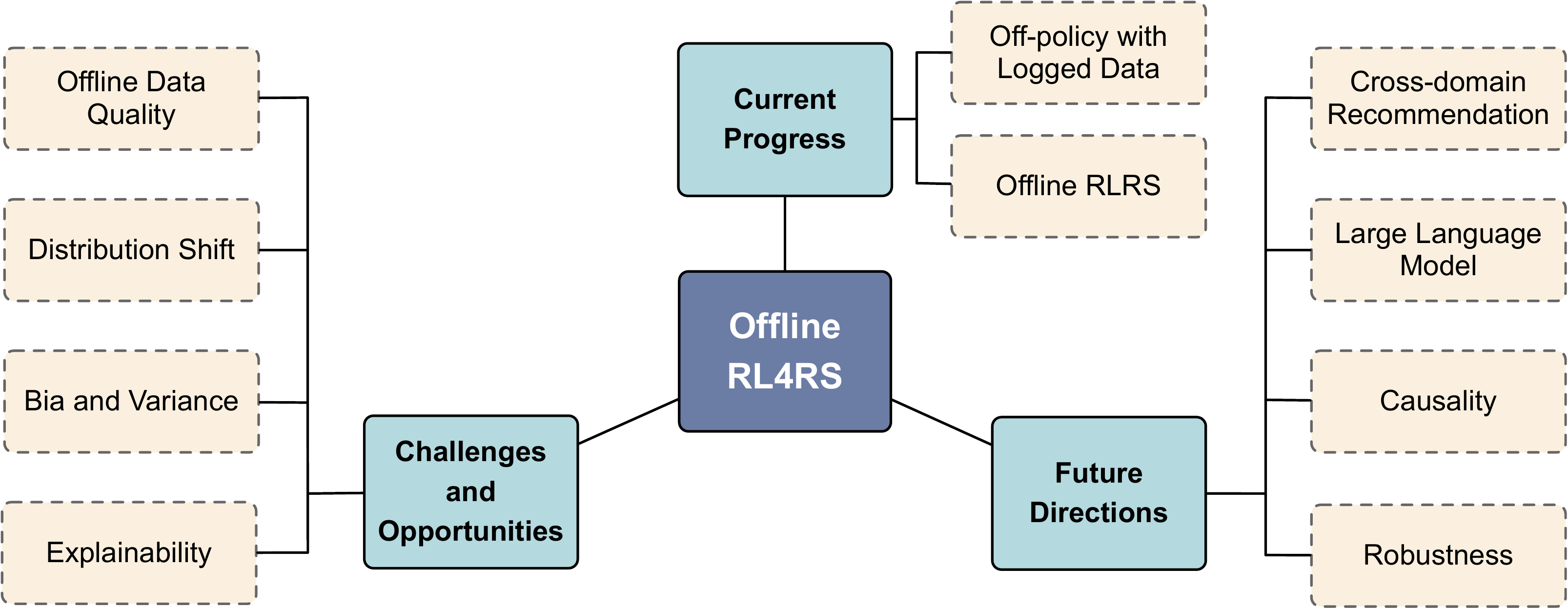}
    \caption{The overall structure of this survey including the section index.}
    \label{fig:tax}
\end{figure}
\subsection{Relations to  existing surveys}
Two existing surveys have centered on the topic of RL in RS~\cite{chen2023deep,afsar2022reinforcement}. While \citet{afsar2022reinforcement} provides an overview of RL in RS, it does not comprehensively explore the expanding realm of deep RL. In contrast, \cite{chen2023deep} delves more deeply into the analysis and discussion of RL in RS, but predominantly focuses on online RL and its RS applications. It's noteworthy that \cite{chen2023deep} identifies offline RL in RS as a potential future direction but does not offer an all-encompassing review of this area. The limited coverage of offline RL in RS can be attributed to its emergence around the same time as these two surveys. Furthermore, due to the recent establishment of the offline RL concept, certain works examined in these two existing surveys are classified as special cases of policy-based methods. Differently, this survey endeavors to refine these categorizations by reclassifying prior works into the domain of offline RL in RS. Additionally, we extend the literature to encompass the most recent developments in offline RL for RS, thereby augmenting our understanding of recent progress in this field.
\subsection{Structure of this Survey}
This survey is structured into four distinct sections. Firstly, we offer an introduction to RL basics, providing readers with a foundational understanding of various RL algorithms, including Q-Learning, Policy-based Methods, Actor-Critic Methods, and Model-based RL. Subsequently, we delve into the concept of offline RL and present a problem formulation that explores how to integrate recommender systems (RS) into the offline RL framework.
Continuing, we conduct a comprehensive review of existing works from two main perspectives: off-policy evaluation using logged data and the realm of offline RL in RS. This examination highlights current research trends and insights. Following the review, we outline the open challenges and promising opportunities that warrant in-depth exploration.
Finally, building upon the identified challenges and opportunities, we propose potential future directions that could serve as solutions to these challenges. This forward-looking section aims to guide future research endeavors in the field, by suggesting pathways to address the outstanding issues and capitalize on the untapped opportunities. 

\section{Offline RL Overview and Problem Statement}
In this section, we delve into fundamental concepts essential to understanding the field of RL. We initiate with RL preliminaries, encompassing Markov Decision Processes, On-Policy and Off-Policy Learning, and Typical RL algorithms. In doing so, we establish the foundational understanding by clarifying key principles and terminologies employed throughout this survey. Subsequently, we shift our focus toward the concept of offline RL and how it can be used to formulate RS. For the sake of clarity, we have summarized the common notations used in this survey in~\Cref{tab:notations}.
\subsection{Markov Decision Process}
\begin{table}[ht]
    \centering
    \caption{Common notations used in this survey}
    \resizebox{\textwidth}{!}{\begin{tabular}{cc|ccc}
        \hline
        Notations & Name & Notations & Name & Notes \\\hline
         $\mathcal{M}$  & Markov Decision Process   &           $s$ &   State     &  User related info    \\
         $\pi_\beta$  &  Behavior Policy    &           $a$ &   Action      &  Recommended item(s)    \\
         $\gamma$ & Discount Factor      &           $\pi$  & Policy  & Recommendation policy       \\
          $\mathbb{E}$ & Expected Value      &          $\mathcal{R}(\cdot,\cdot)$ & Reward function  &  Users' click behavior    \\
          $\theta$ & Policy Parameter      &    $\mathcal{D}$     & Offline Dataset  & Set of $\{(s_t,a_t,s_{t+1},r_t)\}$    \\\hline
    \end{tabular}}
    \label{tab:notations}
\end{table}

In this section, we shall expound upon fundamental concepts within the realm of RL, adhering closely to established standard definitions as outlined in\cite{sutton2018reinforcement}. RL deals with the challenge of learning how to control dynamic systems in a broad context. RL4RS are typically described by fully observed Markov decision processes (MDP) or partially observed ones known as Partially Observable Markov Decision Processes (POMDP). Moreover, we will also provide 
\begin{dfn}[Markov decision process]
    The Markov decision process is formalized as the tuple $\mathcal{M}=\{\mathcal{S},\mathcal{A},\mathcal{P},\mathcal{R},\gamma\}$. Within this structure, each component serves a distinct purpose: $\mathcal{S}$ encompasses the set of states $s\in\mathcal{S}$, capable of adopting discrete or continuous values, potentially even multi-dimensional vectors. $\mathcal{A}$ characterizes the set of actions $a\in\mathcal{A}$, which may be discrete or continuous in nature. $\mathcal{P}$ defines a conditional probability distribution, $\mathcal{P}(s_{t+1}|s_t,a_t)$, delineating the progression of the system's dynamics over time. $\mathcal{R}:\mathcal{S}\times\mathcal{A}\rightarrow \mathbb{R}$ serves as the reward function, linking states and actions to real-valued rewards. $\gamma\in[0,1]$ assumes the role of a scalar discount factor, influencing the extent to which future rewards are taken into consideration.
\end{dfn}
Throughout most of this article, we will primarily employ fully-observed formalism. However, we also include the definition of the partially observed Markov decision process (POMDP) to ensure comprehensiveness. The MDP definition can be extended to the partially observed setting in the following manner:
\begin{dfn}[Partially observed Markov decision process]
    The partially observed Markov decision process is defined as a tuple $\mathcal{M}=\{\mathcal{S},\mathcal{A},\mathcal{O},\mathcal{P},\mathcal{R},\gamma\}$, where $\mathcal{S},\mathcal{A},\mathcal{P},\mathcal{R},\gamma$ are defined as before, $\mathcal{O}$ is a set of observations, where each observation is given by $o\in\mathcal{O}$.
\end{dfn}
The ultimate objective within a RL problem is to acquire a policy, denoted as $\pi$, which establishes a probability distribution over actions conditioned upon states, $\pi(a_t|s_t)$, or alternatively conditioned upon observations within the context of partially observed scenarios, $\pi(a_t|o_t)$. From these definitions, we can derive the trajectory distribution. A trajectory in this context refers to a sequence encompassing both states and actions, spanning a length of $T$, represented as $\tau=\{s_0,a_0,\cdots,s_T,a_T\}$. It is noteworthy that the parameter $T$ can be an infinite value, implying the consideration of scenarios with an indefinite time horizon, as seen in infinite horizon MDP~\cite{sutton2018reinforcement}.

The trajectory distribution $p_\pi$ for a given MDP tuple $\mathcal{M}$ and policy $\pi$ is given by
\begin{align}
    p_\pi(\tau) = d_0(s_0)\prod_{t=0}^T\pi(a_t|s_t)\mathcal{P}(s_{t+1}|s_t,a_t),
\end{align}
where $d_0(s_0)$ represents the initial state distribution.
This definition can easily be extended into the partially observed setting by including the observations $o_t$. The RL objective $J(\pi)$,  can then be written as an expectation under this trajectory distribution:
\begin{align}
    J(\pi) = \mathbb{E}_{\tau\sim p_\pi(\tau)}\bigg[\sum_{t=0}^T\gamma^t \mathcal{R}(s_t,a_t)\bigg]. \label{eq:objective}
\end{align}
\subsection{On-Policy and Off-Policy Learning}
While the process of interaction unfolds, gathering additional episodes enhances the precision of the function estimates. Nevertheless, a challenge arises. If the policy improvement algorithm consistently adjusts the policy in a greedy manner---prioritizing actions with immediate rewards---then actions and states lying outside this advantageous route might not be adequately sampled. Consequently, superior rewards that could exist in these unexplored areas remain concealed from the learning process. Fundamentally, we confront a decision between opting for the optimal choice based on existing data or delving deeper into exploration to collect more data. This predicament is commonly recognized as the Exploration vs.~Exploitation Dilemma.

What we need is a middle ground between these two extremes. Pure exploration would require a significant amount of time to collect the necessary information, while pure exploitation could trap the agent in a local reward maximum. To address this, there are two approaches that ensure all actions are adequately sampled, known as \textit{on-policy} and \textit{off-policy} methods. 

\textit{On-policy} methods resolve the exploration vs.~exploitation dilemma by incorporating randomness through a soft policy. This means that non-greedy actions are chosen with some probability. These policies are referred to as $\epsilon$-greedy policies because they select random actions with a probability of $\epsilon$ and follow the optimal action with a probability of 1-$\epsilon$.

Since the probability of randomly selecting an action from the action space is $\epsilon$, the probability of choosing any specific non-optimal action is $\epsilon/|\mathcal{A}(s)|$. On the other hand, the probability of following the optimal action will always be slightly higher due to the 1 - $\epsilon$ probability of selecting it outright and the $\epsilon/|\mathcal{A}(s)|$ probability of choosing it through sampling the action space:

\textit{Off-policy} methods offer a different solution to the exploration vs.~exploitation problem. While on-policy algorithms attempt to improve the same $\epsilon$-greedy policy used for exploration, off-policy approaches utilize two distinct policies: a behavior policy and a target policy. The behavioral policy (denoted as $\pi_\beta$) is employed for exploration and episode generation, while the target or goal policy (denoted as $\pi$) is used for function estimation and improvement.

The efficacy of this approach lies in the capacity of the target policy $\pi$ to attain a balanced perspective of the environment, enabling it to assimilate insights from the behavioral policy $b$, while concurrently capturing advantageous actions and seeking further improvements. Nevertheless, it is imperative to acknowledge that in off-policy learning, a distributional discrepancy arises between the target policy estimation and the sampled policy. Consequently, a widely employed technique known as importance sampling is applied to address this disparity~\cite{li2017deep}.

\subsection{Typical RL algorithms}
Let's briefly outline various types of RL algorithms and present their definitions. At a high level, all standard RL algorithms follow a common learning loop: the agent engages with the MDP $\mathcal{M}$ using a behavior policy $\pi_\beta$. This behavior policy, which could or could not align with $\pi(a|s)$, leads the agent to observe the current state $s_t$, choose an action $a_t$, and then witness the subsequent state $s_{t+1}$ and the reward value $r_t= \mathcal{R}(s_t,a_t)$. This sequence can repeat over multiple steps, allowing the agent to gather transitions $\{s_t,a_t,s_{t+1},r_t\}$. These observed transitions are then used by the agent to adjust its policy, and this update process might incorporate earlier observed transitions as well. We'll denote the set of available transitions for policy updating as $\mathcal{D}=\{(s_t,a_t,s_{t+1},r_t)\}$. This set could encompass all the transitions seen thus far or a subset thereof.

\textbf{Q-learning}~\cite{watkins1992q} is an off-policy value-based learning scheme aimed at finding a target policy that selects the best action:
\begin{align}
\pi(s) = \arg\max_a Q_\pi (s,a)
\end{align}
Here, $Q_u (s,a)$ represents the Q-value and applies to a discrete action space. For deterministic policies, the Q-value can be computed as:
\begin{align}
Q (s_t,a_t) = \mathbb{E}_{\tau \sim \pi}[r(s_t,a_t) + \gamma Q(s'_t, a'_t)].
\end{align}

Deep Q learning (DQN)~\cite{mnih2015human} employs deep learning to approximate a nonlinear Q function parameterized by $\theta_q$: $Q_{\theta_q} (s,a)$. DQN involves a network $Q_{\theta_q}$ that's updated asynchronously by minimizing the Mean Squared Error (MSE) as defined by:
\begin{align}
\mathcal{L}(\theta_q) = \mathbb{E}{\tau \sim \pi}\Big[Q{\theta_q}(s_t,a_t)-(r(s_t,a_t) + \gamma Q_{\theta_q}(s'{t},a'{t}))\Big]^2 \label{dqnloss}
\end{align}
In this equation, $\tau$ signifies a sampled trajectory including $(s,a,s',r(s,a))$. Notably, $s'_t$ and $a'_t$ originate from the behavior policy $\pi_b$, while $s,a$ come from the target policy $\pi$.

Furthermore, the concept of value functions plays a role. These assess states and actions. The value function $V_\pi(s)$ evaluates states, and $Q_\pi(s_t,a_t)$ evaluates actions. The relationship between them is given by:
\begin{align}
V_\pi(s) = \mathbb{E}{a\sim\pi}[Q\pi(s,a)].
\end{align}

The value function gets updated via the Temporal Difference (TD) method:
\begin{align}
V_\pi(s_t) \leftarrow V_\pi(s_t) + \alpha[r(s'_t,a't) + \gamma V\pi(s't) - V\pi(s_t)],
\end{align}
where $\alpha$ represents the learning rate.

\textbf{Policy gradient}~\cite{williams1992simple} is a technique used in reinforcement learning that tackles scenarios where actions are high-dimensional or continuous—something not easily managed by Q-learning. Unlike Q-learning, which focuses on finding optimal actions, policy gradient aims to find optimal parameters $\theta$ for a policy $\pi_{\theta}$ in order to maximize the total reward.

The central goal of policy gradient is to maximize the expected return, or accumulated reward, starting from the initial state. This is captured by the equation:
\begin{align}
    J(\pi_\theta) = \mathbb{E}_{\tau \sim \pi_{\theta}}[r(\tau)] = \int\pi_{\theta}(\tau) r(\tau)d\tau
\end{align}
Here, $\pi_{\theta}(\tau)$ signifies the probability of observing trajectory $\tau$. The technique learns the optimal parameter $\theta$ by computing the gradient $\nabla_\theta J(\pi_\theta)$ as follows:
\begin{align}
    \nabla_\theta J(\pi_\theta) &= \mathbb{E}_{\tau \sim d_{\pi_\theta}}\left[\sum_{t=1}^Tr(s_t,a_t)\sum_{t=1}^T\nabla_{\theta} \log \pi_{\theta}(s_t,a_t)\right].
\end{align}
In the above equation, $d_{\pi_\theta}$ is the distribution of trajectories generated by policy $\pi_\theta$.

The derivation involves the substitution:
\begin{align}
    \pi_{\theta}(\tau) = p(s_1)\prod_{t=1}^T \pi_{\theta}(s_t,a_t)p(s_{t+1}|s_t,a_t)
\end{align}
Here, $p(\cdot)$ is independent of the policy parameter $\theta$, and for simplicity, it's not explicitly included in the derivation.

Prior policy gradient algorithms, like REINFORCE, have often used Monte-Carlo sampling to estimate $\tau$ from $d_{\pi_{\theta}}$.

\textbf{Actor-critic networks} bring together the strengths of both Q-learning and policy gradient techniques. They can function as either on-policy methods~\cite{konda2000actor} or off-policy methods~\cite{degris2012off}. An actor-critic network is composed of two key components:
\begin{itemize}
    \item The actor: This component optimizes the policy $\pi_\theta$ based on the guidance provided by $\nabla_\theta J(\pi_{\theta})$.
    \item The critic: The critic evaluates the learned policy $\pi_\theta$ using the Q-value function $Q_{\theta_q} (s,a)$.
\end{itemize}
The overall gradient expression for an actor-critic network is as follows:
\begin{align}
\mathbb{E}{s \sim d{\pi_\theta}}[Q_{\theta_q}(s,a)\nabla_{\theta} \log \pi_{\theta}(s,a)].
\end{align}

In the case of off-policy learning, the value function for $\pi_\theta (a|s)$ can be further defined using deterministic policy gradient (DPG):
\begin{align}
\mathbb{E}{s \sim d{\pi_\theta}}[\nabla_a Q_{\theta_q}(s,a)|{a=\pi\theta(s)}\nabla_{\theta} \pi_{\theta}(s,a)].
\end{align}

It's worth noting that while traditional policy gradient calculations involve integrating over both the state space $\mathcal{S}$ and the action space $\mathcal{A}$, DPG only requires integrating over the state space $\mathcal{S}$. In DPG, given a state $s\in\mathcal{S}$, there corresponds only one action $a\in\mathcal{A}$ such that $\mu_\theta(s) = a$.

\textbf{Model-based RL} is a broad term encompassing methods that rely on explicit estimates of the transition or dynamics function $\mathcal{P}$. The distinguishing feature of model-based RL is that it assumes the dynamics model $\mathcal{P}$ is known and can be learned. This is in contrast to other forms of RL where such a dynamics model is neither known nor learnable.

\subsection{Offline RL}
The offline RL problem can be defined as a data-driven formulation of the RL problem~\cite{levine2020offline}. The ultimate objective remains centered on optimizing the goal presented in~\Cref{eq:objective}. Notably, the agent's capacity to engage with the environment and amass supplementary transitions using the behavior policy is nullified. Instead, the learning algorithm receives a fixed collection of transitions denoted as $\mathcal{D}=\{s_t^i,a_t^i,s_{t+1}^i,r_t^i\}$, and its task is to acquire the most optimal policy using this provided dataset. This approach aligns more closely with the supervised learning paradigm, and we can view $\mathcal{D}$ as the training dataset for the policy.

Fundamentally, offline RL necessitates that the learning algorithm comprehends the underlying dynamics of the MDP $\mathcal{M}$ solely from a fixed dataset. Subsequently, it must create a policy $\pi(a|s)$ that achieves the highest cumulative reward during the interaction with the MDP. We will denote the distribution over states and actions in $\mathcal{D}$ as $\pi_\beta$ (also referred to as the behavior policy). Here, we assume that state-action pairs $(s,a)\in\mathcal{D}$ are drawn from $s\sim d^{\pi_{\beta}}(s)$, and actions are sampled according to the behavior policy, i.e., $a\sim\pi_{\beta}(a|s)$.

\begin{figure}[h]
    \centering
    \includegraphics[width=\linewidth]{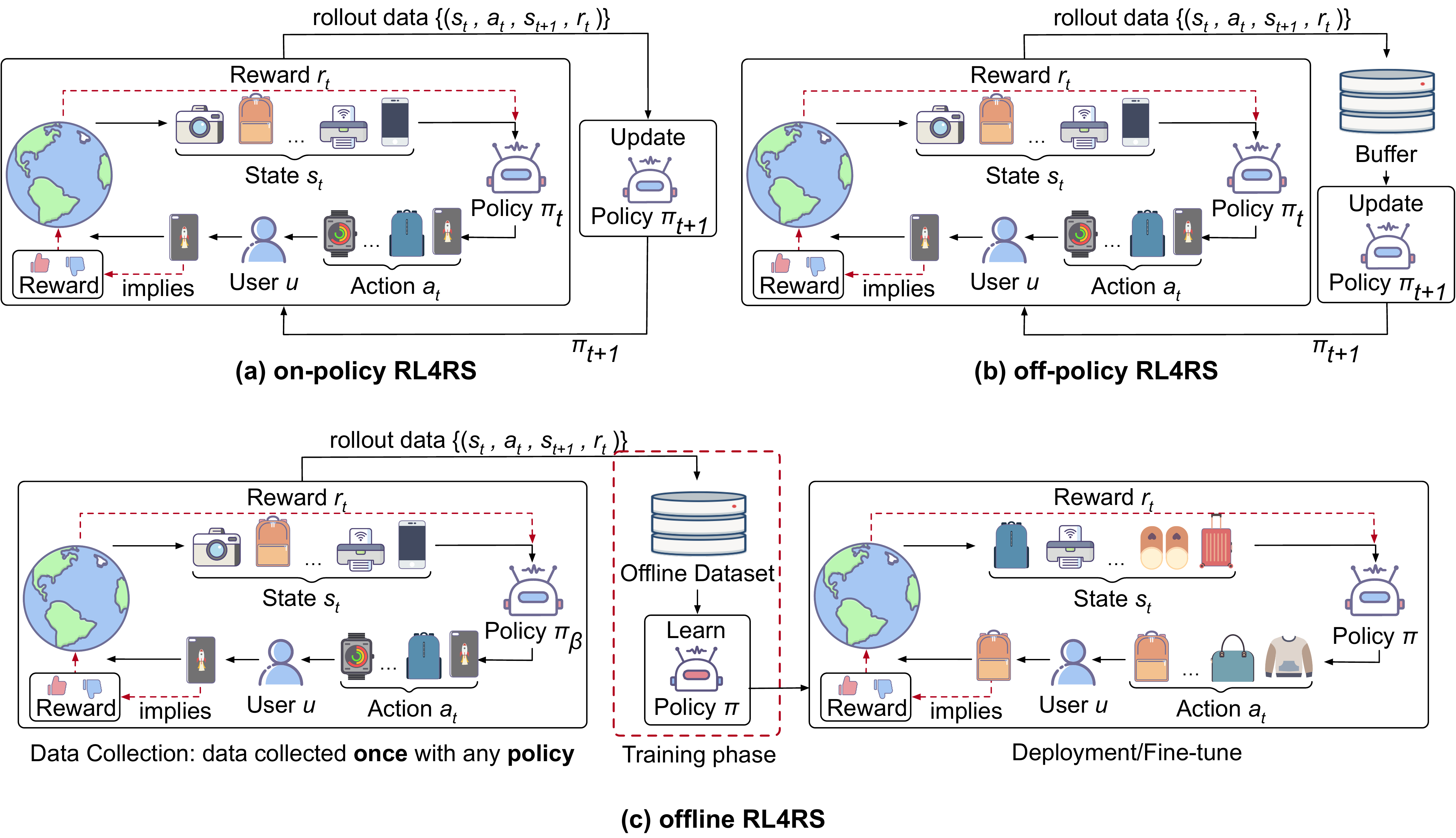}
    \caption{Illustration of classic on-policy RL (a), classic off-policy RL (b), and offline RL (c). Where (a) and (b) can also be recognized as online RL.}
    \label{fig:different-rl}
\end{figure}

This problem formulation has been expressed using a range of terminologies. Within the field of RS, the term that frequently induces confusion is ``off-policy RL''. This phrase is commonly employed as a broad label encompassing all RL algorithms that can leverage datasets of transitions $\mathcal{D}$, wherein the actions in each transition were acquired using policies distinct from the current policy $\pi(a|s)$. However, it's important to note that the term ``off-policy'' typically signifies an RL algorithm where the behavior policy $\pi_\beta$ differs from the target policy $\pi$, as discussed earlier. This distinction can sometimes cause confusion. Hence, the terms ``fully off-policy RL'' or ``offline RL'' are recently introduced to indicate situations where no additional online data collection takes place. We have presented various illustrations of distinct RL approaches to emphasize the disparities between them in~\Cref{fig:different-rl}.

The challenge of offline RL can be tackled through algorithms belonging to any of the four main categories in RL: Q-learning, policy gradient, actor-critic, and model-based RL. In principle, any off-policy RL algorithm could function as an offline RL approach when the online interaction process is excluded. For instance, a straightforward offline RL technique can be created by employing Q-learning \textit{without requiring supplementary online exploration}. This method utilizes the dataset $\mathcal{D}$ to pre-fill the data buffer.

\subsection{Offline RL4RS - Problem Formulation}
In this section, we establish a problem formulation for Offline RL4RS. We begin with a standard MDP framework, commonly used in RS. The setup involves a set of users denoted as $\mathcal{U} = {u, u_1, u_2, u_3, ...}$ and a set of items denoted as $\mathcal{I} = {i, i_1, i_2, i_3, ...}$. The process begins with the system recommending item $i$ to user $u$ and receiving feedback $f_i^u$. This feedback is then utilized to enhance future recommendations, leading to the identification of an optimal policy $\pi^*$ that guides the selection of items to recommend in order to achieve positive feedback.

The MDP framework treats the user as the environment while the system acts as the agent. The fundamental components within the MDP context, especially in Deep Reinforcement Learning (DRL)-based RS, include:
\begin{itemize}
    \item  State $\mathcal{S}$: At a given time $t$, the state $s_t\in\mathcal{S}$ is defined by a combination of the user's characteristics and the recent $l$ items that the user has shown interest in prior to time $t$. This may also include demographic information if relevant.

    \item Action $\mathcal{A}$: The action $a_t \in\mathcal{A}$ represents the agent's prediction of the user's evolving preferences at time $t$. Here, $\mathcal{A}$ encompasses the entire set of potential candidate items, which could be vast, potentially numbering in the millions.

    \item Transition Probability $\mathcal{P}$: The transition probability $p(s_{t+1}|s_t,a_t)$ quantifies the likelihood of transitioning from state $s_t$ to $s_{t+1}$ when the agent performs action $a_t$. In the context of a recommender system, this probability corresponds to the likelihood of user behavior changes.

    \item Reward function $\mathcal{R}$: After the agent selects an appropriate action $a_t$ based on the current state $s_t$ at time $t$, the user receives the item recommended by the agent. The feedback from the user regarding the recommended item contributes to the reward $r_t=\mathcal{R}(s_t,a_t)$. This reward reflects the user's response and guides the enhancement of the learned policy $\pi$ by the recommendation agent.

    \item Discount Factor $\gamma$: The discount factor $\gamma \in [0,1]$ is employed to balance the consideration of future and immediate rewards. A value of $\gamma=0$ indicates the agent prioritizes immediate rewards, while a non-zero value implies a blend of both immediate and future rewards.

    \item Offline Dataset $\mathcal{D}$: The offline dataset $\mathcal{D}$ is amassed by an unknown behavior policy $\pi_\beta$. This dataset serves as the historical records of user interactions and is utilized to improve the recommendation policy.
\end{itemize}
This MDP-based framework lays the groundwork for Offline RL4RS, where the aim is to devise effective recommendation policies using historical interaction data, even when the data is collected under an unknown or different behavior policy. If a POMDP is used, we just need to add the observation $\mathcal{O}$ which is the partial information from users and $l$ items in which the user was interested before time $t$.
\begin{dfn}
Given an offline dataset $\mathcal{D}$, which contains the trajectories when user $u\in\mathcal{U}$ interacts with the system for a certain period with an unknown behavior policy $\pi_\beta$, the RL agent aims to learn a policy $\pi$ from the offline dataset $\mathcal{D}$. After that, the trained policy $\pi$ will be deployed/tested on a production or evaluation environment with a similar scenario with the collected dataset $\mathcal{D}$.
\end{dfn}

\section{Current Progress in Offline RL4RS}
In this section, we survey offline RL-based RS. Generally speaking, it can be divided into two categories: off-policy with logged data (i.e., ``full'' off-policy) and offline RL. These two concepts are generally the same except for some specific settings in off-policy methods such as assuming bandit conditions. Due to the recent introduction of offline RL, we have opted to distinguish and separate these for clarity and to prevent potential confusion.

\subsection{Off-policy with Logged Data}
\subsubsection{Off-policy Evaluation} The typical method in this domain is known as off-policy evaluation. 
Off-policy evaluation methods are rooted in the direct estimation of policy returns. These methods often utilize a technique known as importance sampling, which involves estimating the return of a given policy or approximating the corresponding policy gradient. A straightforward application of importance sampling involves using trajectories sampled from $\pi_\beta(\tau)$ to derive an unbiased estimator of $J(\pi)$:
\begin{align}
    J(\pi_\theta)= \mathbb{E}_{\tau\sim\pi_{\beta(\tau)}}\bigg[\frac{\pi_\theta(\tau)}{\pi_\beta(\tau)}\sum_{t=0}^T \gamma^t \mathcal{R}(s,a)\bigg] & =  \mathbb{E}_{\tau\sim\pi_{\beta(\tau)}}\bigg[\prod_{t=0}^T\frac{\pi_\theta(a_s|s_t)}{\pi_\beta(s_t|a_t)}\sum_{t=0}^T \gamma^t \mathcal{R}(s,a)\bigg] \\
    & \approx \sum_{i=1}^n w_T^i \sum_{t=0}^T \gamma^t r^i_t. \label{eq:important_sampling}
\end{align}
However, this estimator often exhibits high variance, particularly if $T$ (the time horizon) is large, due to the product of importance weights. To address this, a weighted importance sampling estimator can be used. This involves dividing the weights by $\sum_{i=1}^n w_T^i$ to normalize them, resulting in a biased estimator with significantly lower variance, while still maintaining strong consistency.

When considering the estimation of Q-values for each state-action pair $(s_t,a_t)$, denoted as $\hat{Q}^\pi(s_t,a_t)$, an approximate model comes into play. This model could be derived from estimating the transition probability $\mathcal{P}(s_{t+1}|s_t,a_t)$ of the Markov Decision Process (MDP) and subsequently solving for the corresponding Q-function. Alternatively, other methods for approximating Q-values could be employed.

The integration of these approximated Q-values as control variates within the importance sampled estimator leads to an enhanced approach:
\begin{align}
J(\pi_\theta) = \sum_{i=1}^n\sum_{t=0}^T \gamma^t \Big(w_t^i(r_t^i-\hat{Q}^\pi(s_t,a_t)) - w_{t-1}^i\mathbb{E}{a\sim\pi\theta(a|s_t)}[\hat{Q}^\pi(s_t,a)]\Big).
\end{align}
This method, referred to as a doubly robust estimator \cite{jiang2016doubly} , exhibits unbiasedness either when $\pi_\beta$ is known or when the model is accurate. In essence, it leverages both the unbiasedness of the importance sampling method and the approximated Q-values to produce an estimator with lower variance and strong consistency.

\subsubsection{Recent works}\label{sec:off-policy-evaluation}
\begin{figure}[h]
    \centering
    \includegraphics[width=\linewidth]{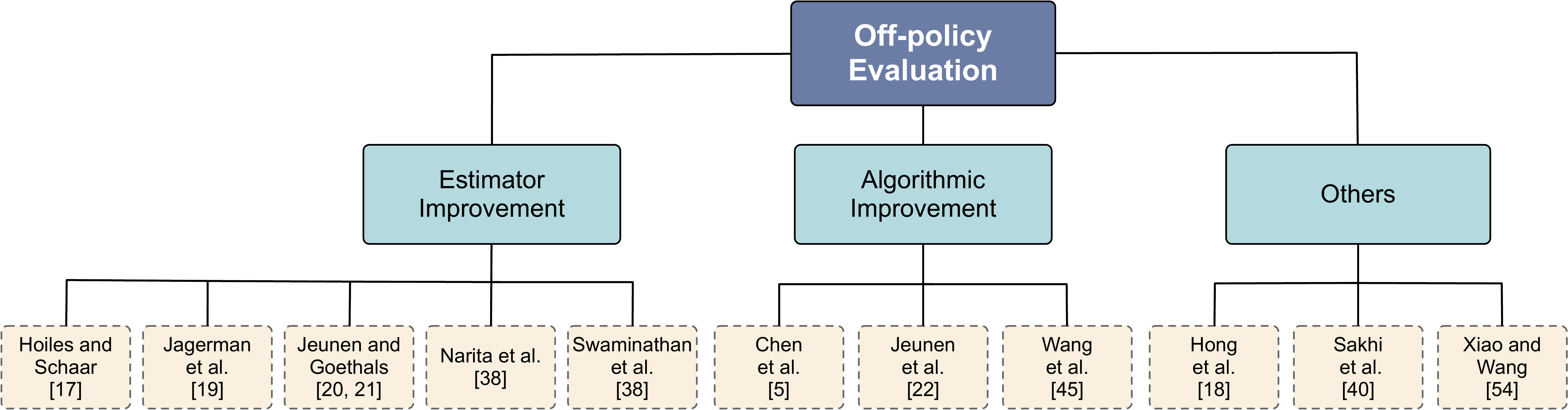}
    \caption{Off-Policy Evaluation Works Classifications}
    \label{fig:off-policy-ev-class}
\end{figure}

The recent advancements in off-policy using logged data method can be broadly categorized into three distinct domains: estimator improvement (focus on the discrepancy between the offline data and online data), algorithmic improvement (focus on the recommendation algorithm itself), and miscellaneous application domains. We have compiled a summary of these works in \Cref{fig:off-policy-ev-class}.

\textbf{Estimator Improvement}
\citet{hoiles2016bounded} focus on the problem of student course scheduling and curriculum design. It proposes an algorithm for personalized course recommendation and curriculum design based on logged student data. The algorithm uses a regression estimator for contextual multi-armed bandits and provides guarantees on their predictive performance. The paper also addresses the issue of missing data and provides guidelines for including expert domain knowledge in the recommendations. The algorithms can be used to identify curriculum gaps and provide recommendations for course schedules. The paper also discusses off-policy evaluation techniques and the use of the regression estimator for estimating the expected reward of a new policy. One drawback is that the proposed approach assumes a fixed set of courses and does not consider the dynamic nature of course offerings and student preferences.

\citet{swaminathan2017off} address the problem of off-policy evaluation and optimization in combinatorial contextual bandits. The motivation behind this research is the need to estimate the reward of a new target policy using data collected by a different logging policy. The authors propose a pseudoinverse (PI) estimator that makes a linearity assumption about the evaluated metric, allowing for more efficient estimation compared to importance sampling. The PI estimator requires exponentially fewer samples to achieve a given error bound and can be used for off-policy optimization as well.
The methodology involves using the PI estimator to impute action-level rewards for each context, enabling direct optimization of whole-page metrics through pointwise learning to rank algorithms. The authors demonstrate the effectiveness of their approach on real-world search ranking datasets, showing that the PI estimator outperforms prior baselines in terms of off-policy evaluation of whole-page metrics.
This method has several limitations. One drawback of this method is that it relies on the linearity assumption, which may not always hold in practice. Moreover, there is a bias-variance tradeoff between the weighted pseudoinverse (wPI) method and the direct method, with wPI showing bias for the Expected Reciprocal Rank metric. The wPI method also deteriorates for larger slate spaces and is sensitive to linearity assumptions. These drawbacks highlight areas where further refinement and research are needed to enhance the robustness and effectiveness of the approach.

\citet{jeunen2021pessimistic,jeunen2023pessimistic} focus on improving the recommendation performance of policies that rely on value-based models (i.e., Q-learning) of expected reward. The authors propose a pessimistic reward modeling framework that incorporates Bayesian uncertainty estimates to express skepticism about the reward model. This allows for the generation of conservative decision rules based on lower-confidence-bound estimates, rather than the usual maximum likelihood or maximum PI estimates. The approach is agnostic to the logging policy and does not require propensity scores, making it more flexible and avoiding the limitations of inverse propensity score weighting.
The methodology involves training reward models using a range of datasets generated under different environmental conditions. The authors compare the performance of policies that act based on reward models using maximum likelihood or maximum PI estimates, with policies that use lower confidence bounds based on tuned parameters. The evaluation is done through simulated A/B tests, with the resulting click-through-rate (CTR) estimates compared to the logging policy and an unattainable skyline policy. The experiments show that the pessimistic decision-making approach consistently decreases post-decision disappointment and can significantly increase the policy's attained CTR.
One drawback of this approach is that it relies on the assumption that the reward estimates are conditionally unbiased, which may not always hold in practice. The authors acknowledge that underfitting and model misspecification can make this assumption unrealistic. Additionally, the approach requires tuning the hyperparameter alpha, which determines the lower confidence bound, and finding the optimal value may not always be straightforward.

\citet{narita2021debiased} proposes a new off-policy evaluation method for RL4RS. The motivation behind this work is to address the limitations of existing estimators, such as inverse propensity weighting and doubly robust estimators, which suffer from bias and overfitting issues. The authors introduce a new estimator that combines the doubly robust estimator with double/debiased machine learning techniques. The key features of this estimator are its robustness to bias in behavior policy and state-action value function estimates, as well as the use of a sample-splitting procedure called cross-fitting to remove overfitting bias. However, the experiments are limited to specific domains, such as the CartPole-v0 environment and online ads, and it is unclear how the estimator would perform in other tasks in RS.

\citet{jagerman2019people} address the problem of off-policy evaluation in non-stationary environments, where user preferences change over time. Existing off-policy evaluation techniques fail to work in such environments. It proposes several off-policy estimators that operate well in non-stationary environments. These estimators rely more on recent bandit feedback and accurately capture changes in user preferences. They provide a rigorous analysis of the proposed estimators' bias and show that the bias does not grow over time, unlike the standard Inverse Propensity Scoring (IPS) estimator. They also create adaptive variants of the estimators that change their parameters in real-time to improve estimation performance. Extensive empirical evaluation on recommendation datasets shows that the proposed estimators significantly outperform the regular IPS estimator and provide a more accurate estimation of a policy's true performance. One drawback of the work is the trade-off between bias and variance. While the estimators avoid a bias term that grows with time, they introduce variance that scales with the window size or decay factor. Choosing a smaller window size or larger decay factor reduces bias but increases variance, and vice versa. Finding the optimal balance between bias and variance is a challenge. 


\textbf{Algorithmic Improvement}
\citet{wang2020off} address the problem of designing a stable off-policy RL method for RS. 
Moreover, the exploration error is also highlighted, which arises from the mismatch between the recommendation policy and the distribution of customers' feedback in the training data. 
This exploration error can lead to unstable training processes and potentially diverging results. To mitigate this problem, the authors propose an off-policy logged data method called Generator Constrained deep Q-learning (GCQ). GCQ combines a neural generator that simulates customers' possible feedback with a Q-network that selects the highest valued action to form the recommendation policy. The authors also design the generator's architecture based on Huffman Trees to reduce decision time. One drawback of this work is the limited capability to handle long sequences of user behavior.

\citet{chen2019top} address the problem of data biases that arise when applying policy gradient methods in a recommendation system. The primary goal is to address the distribution mismatch from the behavior policy $\pi_\beta$ and the learned policy $\pi$. As a result, an off-policy-corrected gradient estimator is introduced to reduce the variance of each gradient term while still correcting for the bias of a non-corrected policy gradient. A recurrent neural network (RNN) is adopted to model the user state at each time step. To estimate the behavior policy $\pi_\beta$, which is a mixture of the policies of multiple agents in the system, the authors use a context-dependent neural estimator which is a contextual bandit based method. 
One drawback of the proposed method is the variance of the estimator, which can be large when there are very low or high values of the importance weights. To reduce this variance, the authors take a first-order approximation and ignore the state visitation differences under the two policies. This results in a slightly biased estimator with a lower variance. 
Another drawback is the difficulty in estimating the behavior policy $\pi_beta$, especially when there are multiple agents in the system and the collected trajectories are generated by a mixture of deterministic policies and stochastic policies. 

\citet{jeunen2020joint} propose a new approach called the Dual Bandit, which combines value-based and policy-based methods to improve performance in recommendation settings. It highlights that existing offline evaluation results are often even contradictory over different runs and datasets, or extremely hard to reproduce in a robust manner. Hence, they introduce simulation environments as an alternative and reproducible evaluation approach.

\textbf{Others}
\citet{sakhi2020blob} introduce a probabilistic model known as BLOB (Bayesian Latent Organic Bandit) designed for bandit-based RS. BLOB aims to enhance recommendation quality by combining organic user behavior (items viewed without intervention) with bandit signals (recommendations and their outcomes). Traditional recommendation algorithms often focus on either organic-based or bandit-based approaches, but the authors recognize the potential to enhance recommendation quality by integrating both aspects. The goal is to create a model that leverages the relationship between organic and bandit behaviors to provide more accurate and personalized recommendations. The proposed model uses a matrix variate prior distribution to relate these two types of behaviors, and variational autoencoders are employed for training. However, the proposed model requires a two-state training process which needs to train the model for organic behavior and bandit signals separately instead of training simultaneously.

\citet{xiao2022towards} present a value ranking algorithm that combines RL and ranking metrics to improve the effectiveness of ranking algorithms. The proposed method uses the concept of extrapolation and regularization to address the challenges of partial and sparse rewards. Extrapolation is used to estimate rewards from logged feedback, while regularization is used to incorporate ranking signals into the RL policy. The authors propose a sequential Expectation-Maximization (EM) framework that alternates between the E-step, which estimates rewards and ranking signals, and the M-step, which optimizes the RL policy. They show that this framework can effectively learn from rewards and ranking signals. This proposed algorithm's drawback lies in the bandit setting, as it doesn't account for future rewards. Additionally, in the full RL setting, it might suffer from the curse of dimensionality.

\citet{hong2023multi} address the complex issue of multi-task off-policy learning from bandit feedback, a challenge that has significant implications for various applications, including RS. It is motivated to develop a solution that can efficiently handle multiple tasks simultaneously, leveraging the relationships between tasks to enhance performance. It proposes a hierarchical off-policy optimization algorithm (HierOPO) to tackle this problem. The problem is formulated as a contextual off-policy optimization within a hierarchical graphical model, specifically focusing on linear Gaussian models. The authors provide an efficient implementation and analysis, proving per-task bounds on the sub-optimality of the learned policies. They demonstrate that using the hierarchy improves performance compared to solving each task independently. The algorithm is evaluated on synthetic problems and applied to a multi-user recommendation system. However, the proposed method is a model-based off-policy approach, the model-based approaches
tend to be biased, due to using a potentially misspecified model. 

\subsection{Offline RL4RS}
In this section, we will provide reviews of existing offline RL4RS methods. Different from off-policy evaluation, offline RL4RS does not limit the setting to bandit-based methods. Moreover, in this part, we have included the off-policy learning based methods as offline RL.
However, the existing works in this field lack organization, with no apparent interconnection among the various works that often emphasize different aspects. Currently, we lack a systematic approach to review these works, resorting to a sequential examination of each one individually.

\citet{ma2020off} discuss off-policy learning in two-stage RS. The proposed method consists of a candidate generation model in the first stage and a ranking model in the second stage. The authors propose a two-stage off-policy policy gradient method that takes into account the ranking model when training the candidate generation model. The proposed method employs IPS to correct the bias and design variance reduction tricks to reduce the variance. However, the proposed method does not provide a comprehensive experiment about how the ranking model and the candidate generation model affect the final performance.

\citet{chen2022off} focus on scaling an off-policy actor-critic algorithm for industrial recommendation systems. The motivation behind their research is to address the challenges of offline evaluation and learning in RS, where only partial feedback is available. The authors propose an approach that combines off-policy learning with importance weighting to estimate the value of state-action pairs under the target policy. They use a critic network to estimate the value function and update the policy network accordingly. The methodology involves minimizing the temporal difference loss and using a Huber loss to handle outliers. The authors also investigate the impact of different estimation methods for the target value function. However, the proposed methods have several limitations. One drawback is the potential bias introduced by using the cumulative future return on the behavior trajectory while ignoring the importance weighting on future trajectories. Another drawback is the conservative nature of the learned policy when using sampling from the learned policy. The softmax policy parameterization used in the approach leads to a more myopic policy, recommending more popular and longer content and less novel content. 

\citet{gao2023alleviating} centre around the problem of the Matthew effect in offline RL based RS. The Matthew effect~\cite{merton1988matthew} describes a phenomenon where popular items or categories are recommended more frequently, leading to the neglect of less popular ones. This bias towards popular items can reduce the diversity in recommendations and decrease user satisfaction. To address the Matthew effect, the authors propose a Debiased model-based Offline RL (DORL) method. DORL introduces a penalty term to the RL algorithm, encouraging exploration and diversity in recommendations. By adding this penalty, the method aims to reduce the bias towards popular items and promote a more varied selection. 

\citet{10.1145/3539618.3591648} address the challenges inherent in designing reward functions and handling large-scale datasets within RL4RS. Traditional RL4RS approaches may fall short in accurately estimating rewards only based on limited observations. To address this problem, a Causal Decision Transformer for RS (CDT4Rec) is proposed, a novel model that integrates offline RL and transformer architecture. CDT4Rec employs a causal mechanism to estimate rewards based on user behavior, allowing for a more accurate understanding of user preferences. The transformer architecture is used to process large datasets and capture dependencies, enabling the model to handle complex data structures. 

\citet{yuan2022offline} is motivated by the challenges associated with optimizing mobile notification systems. Traditional response-prediction models often struggle to accurately attribute the impact to a single notification, leading to inefficiencies in managing and delivering notifications. Recognizing this limitation, the authors aim to explore the application of RL to enhance the decision-making process for sequential notifications, seeking to provide a more effective and targeted approach to mobile notification systems. Hence, an offline RL framework specifically designed for sequential notification decisions is proposed. They introduce a state-marginalized importance sampling policy evaluation approach, which is a novel method to assess the effectiveness of different notification strategies. Through simulations, the authors demonstrate the performance of the approach, and they also present a real-world application of the framework, detailing the practical considerations and results. 

\citet{wang2020offline} are motivated by the challenge of adapting to new users in recommendation systems, particularly when there are limited interactions to understand user preferences. This situation, often referred to as the ``cold-start'' problem, can hinder the ability to provide personalized recommendations that align with long-term user interests. The proposed approach introduces a user context variable to represent user preferences, employing a meta-level model-based RL method for rapid user adaptation. The user model and recommendation agent interact alternately, with the interaction relationship modeled from an information-theoretic perspective.

\citet{zhang2022text} discuss the problem of interactive recommendation with natural-language feedback and proposes an offline RL framework to address the challenges of collecting experience through user interaction. The authors develop a behavior-agnostic off-policy correction framework that leverages the conservative Q-function for off-policy evaluation. This allows for learning effective policies from fixed datasets without further interactions.

\citet{xiao2021general} propose a general offline RL framework for  the interactive recommendation. The proposed method introduces different techniques such as support constraints, supervised regularization, policy constraints, dual constraints, and reward extrapolation. These methods aim to minimize the mismatch between the recommendation policy and logging policy and to balance the supervised signal and task reward. 
\section{Challenges and Opportunities}
Offline RL4RS is an emerging domain that introduces multiple challenges demanding comprehensive exploration. In this section, we aim to outline the open challenges in offline RL4RS. Given that RS fall under the application scope of offline RL, several shared challenges naturally arise. We will begin by addressing some common challenges before delving into the specific challenges unique to RS when utilizing offline RL techniques.

\subsection{High-quality Offline Data and Cold-Start Problems}
One of the most prominent challenges in offline Reinforcement Learning (RL) lies in the fact that the learning process hinges solely on the provided static dataset $\mathcal{D}$. This limitation results in a significant obstacle to enhancing exploration, as exploration falls outside the algorithm's purview. Consequently, if the dataset $\mathcal{D}$ lacks transitions that demonstrate regions of the state space yielding high rewards, the algorithm may be fundamentally incapable of uncovering these rewarding regions.
In contrast to control tasks, which are common in offline RL applications and often face challenges in gathering comprehensive data to facilitate effective learning from high-reward scenarios, the landscape changes when it comes to RS. In this domain, a plethora of offline datasets, such as those from MovieLens, GoodReads, and Amazon, are publicly available. These datasets stem from real-world interactions and adeptly capture users' preferences.

However, RS diverge from traditional offline RL application domains due to their distinct characteristics. To illustrate, let's consider implicit feedback, particularly review data. This kind of data poses a challenge when attempting to embed it within the state space due to its reliance on text. Although techniques like word2vec~\cite{mikolov2013efficient} exist to transform textual data into vectors that might potentially be integrated into the state space, the question of how to effectively guide the agent in utilizing such data in RS remains unexplored.

Another intriguing aspect is the presence of graph data, extensively used in RS to represent social connections, item relationships, and more. The prevalent form of representation is a knowledge graph, which can be transformed into embeddings through the application of Graph Neural Networks (GNN)~\cite{wu2022graph}. Nonetheless, it faces a similar challenge as textual data: how to empower the agent to effectively utilize this information. There are some works investigating graph RL which may be able to provide some directions to offline RL4RS~\cite{xiong2017deeppath,jiang2018graph,madjiheurem2019representation}.

However, a challenge surfaces due to what's known as the ``data sparsity problem''. This means that despite having ample data, there's no assurance that the collected user interactions or behaviors cover all the situations where users have expressed positive feedback, like giving high ratings. In other words, there might be important scenarios where users found something valuable, but the data doesn't reflect those instances well~\cite{chen2022locality}. 

On the other hand, there is s another widely recognized hurdle in RS that also applies to Offline RL4RS: the cold-start problem. Unlike data sparsity, cold-start challenges emerge when the agent aims to provide recommendations to a new user. This issue arises due to the absence of adequate historical data or interactions, which in turn hampers the understanding of preferences and traits related to these new users or items. While addressing the cold-start problem is an ongoing research avenue in conventional RS tasks, it hasn't received sufficient attention in the context of RL4RS. Considering the interactive procedure of the RL4RS, new users have limited contextual information that they can use to formulate the state representation; this contributes to the difficulty of making recommendations. This predicament continues to remain an unsolved puzzle within the realm of offline RL4RS. 

\subsection{Distribution Shift}
A challenge of significant intricacy within the context of offline RL pertains to the effective formulation and addressing of counterfactual queries—a task that might not be readily apparent but is of great importance.

Counterfactual queries, in essence, are defined as hypothetical ``what if'' scenarios. These queries involve creating educated guesses about potential outcomes if the agent were to undertake actions different from those observed in the data. It is the core behind offline RL, as our objective is to learn a policy that can perform better than the behavior recorded in the dataset $\mathcal{D}$. Hence, the agent must execute an action that is different from the learned policy. This situation, unfortunately, places a substantial strain on the capabilities of several prevailing deep-learning methods. Existing methods have been methodically fashioned under the assumption that the data is independence and identical distribution (i.i.d.). In traditional supervised learning based RS, the goal is to train a model to achieve superior performance, such as higher accuracy, recall or precision. The evaluation dataset follows the same distribution as the training dataset. 
Hence, in offline RL4RS, the key point is to learn a policy that can recommend different items (ideally with better feedback) from the behavior recorded in the dataset $\mathcal{D}$.

The challenge behind counterfactual queries is that of distribution shift. The policy is trained under one distribution, but it will be evaluated on a different distribution. Given that such a problem is not widely discussed in the RS literature, we will provide some algorithmic insights from the offline RL perspective to help address this in offline RL4RS.
Distribution shift issues can be addressed in several ways, with the simplest one being to constrain something about the learning process such that the distribution shift is bounded. For example, we can constrain how much the learned policy $\pi(a|s)$ differs from behavior policy $\pi_\beta(a|s)$ by using some techniques like Trust Region Policy Optimization (TRPO)~\cite{schulman2015trust}.

However, if there is a significant disparity between the distribution of the training dataset and that of the evaluation environment, it might lead to the emergence of out-of-distribution (OOD) behavior. Several recent studies have delved into OOD recommendation~\cite{wang2022causal,he2022causpref}, taking into account shifts in user features. These efforts can be categorized into two main groups: OOD generalization~\cite{he2022causpref} and OOD adaptation~\cite{wang2022causal}. 

The underlying notion here is to acquire a causal representation of users' preferences by leveraging their most recent behaviors. This representation is then utilized within a causal graph framework to comprehend how shifts in features could impact users' preferences. Furthermore, the current methodologies primarily target sequential recommendation systems, which share certain properties with MDPs, rendering them relevant to offline RL4RS.

However, this domain is still in its exploratory phase, and it has not garnered substantial attention. As a result, this presents an open challenge with significant potential for further exploration.

\subsection{Bias and Variance Trade-off}
Another prevalent issue within offline RL4RS pertains to the bias inherited from RS, a topic that has recently gained increasing research attention. This bias stems from the nature of offline data, with recent studies~\cite{chen2023bias} revealing that user behavior data are not experimental but rather observational, introducing bias-related challenges.

The prevalence of bias can be attributed to two primary factors. Firstly, the inherent character of user behavior data is observational rather than experimental. In simpler terms, the data fed into RS are susceptible to selection bias. For instance, in a video recommendation system, users tend to engage with, rate, and comment on movies that align with their personal interests. Secondly, a discrepancy in distribution exists, signifying that the distributions of users and items within the recommender system are uneven. This imbalance can lead to a ``popularity bias'', where popular items receive disproportionately frequent recommendations compared to others. Nonetheless, disregarding products within the "long tail" of less popular items can have adverse effects on businesses, given that these items are equally essential, albeit less likely to be discovered by chance.

As mentioned earlier, a substantial portion of existing offline off-policy with logged data methods primarily focus on off-policy evaluation. This approach employs importance sampling to tackle the bias issue. However, the importance sampling gives rise to another hurdle---high variance. While importance sampling already contends with high variance, this issue is further exacerbated in the context of sequential scenarios. In this setting, the importance weights at consecutive time steps are multiplied together (as depicted in Equation \ref{eq:important_sampling}), leading to an exponential amplification of variance.

Approximate and marginalized importance sampling methods mitigate this concern to some extent by circumventing the multiplication of importance weights across multiple time steps. Yet, the fundamental challenge persists: when the behavior policy $\pi_\beta$ substantially diverges from the current learned policy $\pi_\theta$, the importance weights degenerate. Consequently, any estimations of the return or gradient encounter excessive variance, particularly in scenarios characterized by high-dimensional state and action spaces or extended time horizons (as seen in problems like recommendation systems).

For this reason, importance-sampled estimators are most effective when the policy's deviation from the behavior policy remains within a reasonable limit. In the general off-policy setting, this condition generally holds true, as new trajectories are frequently amassed and integrated into the dataset using the latest policy. However, in the offline context, this is not typically the case. Consequently, the extent of enhancement achievable through importance sampling is confined by several factors: (i) the relative suboptimality of the behavior policy; (ii) the dimensionality of the state and action space; (iii) the effective task horizon.

Hence, the tradeoff between bias and variance in offline RL4RS presents an intriguing potential avenue for advancement.
\subsection{Explainability} 
While deep learning-based models can significantly enhance the performance of RS, they often lack interpretability. Consequently, the task of rendering recommender outputs understandable becomes vital, all while maintaining high-quality recommendations. Elevating explainability in RS carries benefits beyond aiding end-users in comprehending suggested items. It empowers system designers to scrutinize the inner workings of RS~\cite{zhang2020explainable}. Additionally, the realm of explainability in RL (RL) has been garnering attention~\cite{heuillet2021explainability}, although the current focus primarily revolves around visualizing learned representations. What remains is an explanation of how the learned policy translates into actionable decisions. In the transition to RL4RS, the emphasis on explainability will shift towards elucidating how the agent justifies its recommended items. Hence, explainability becomes a relatively easy task compared with interpreting the learning process or decision process.

Attention models have emerged as powerful tools that not only bolster predictive performance but also enhance explainability~\cite{zhang2019deep}. For instance,\citet{wang2018reinforcement} introduce an RL framework coupled with an attention model for explainable recommendations. This approach ensures model-agnostic by segregating the recommendation model from the explanation generator. Agents instantiated through attention-based neural networks facilitate the generation of sentence-level explanations. This approach could prove promising given the close connection between offline RL4RS and online RL4RS.

Moreover, with access to offline datasets in offline RL4RS, more solutions become feasible. Knowledge graphs, for instance, contain abundant user and item information, enabling the creation of more personalized, intuitive explanations for recommendation systems~\cite{zhang2020explainable}. However, the processing of graph data presents challenges. One potential strategy involves embedding a pre-learned knowledge graph from the offline dataset into the environment. The final objective then shifts from recommending items to navigating the knowledge graph. As an example, \citet{zhao2020leveraging} extract informative path demonstrations with minimal labeling effort. Then an adversarial actor-critic model for demonstration-guided pathfinding is proposed. This approach enhances recommendation accuracy and explainability through RL and knowledge graph reasoning and can be further expanded by integrating offline RL features.

\section{Future Directions}
In offline RL4RS, several key areas emerge as promising avenues. Cross-domain recommendation systems offer the potential in transferring insights between diverse domains, enhancing recommendation effectiveness. The integration of large language models holds the prospect of enriching contextual understanding and refining user-item interactions. Incorporating causality into offline RL4RS can deepen comprehension of user behaviors, leading to more accurate and interpretable recommendations. The exploration of self-supervised learning and graph-based techniques presents innovative possibilities for capturing intricate user-item relationships. Moreover, addressing uncertainty and fortifying the robustness of RL4RS against noise and adversarial inputs stand out as essential directions for ensuring dependable and consistent recommendation outcomes.
\subsection{Cross-Domain Recommendation}
Cross-domain recommendation refers to the task of providing recommendations to users by leveraging data and knowledge from multiple distinct domains. Cross-domain recommendation systems can be particularly useful in scenarios where user data is sparse within a single domain but might be enriched when multiple domains are combined. Additionally, they enable more comprehensive and diverse recommendations by tapping into different aspects of users' interests. From this viewpoint, we may be able to treat  offline RL4RS as a type of cross-domain recommendation in certain situations. For example, when the evaluation environment is significantly different from the offline dataset $\mathcal{D}$, we may treat the evaluation platform as a new domain and we would like to transfer those learned knowledge from $\mathcal{D}$ into such a platform.  

The challenge in cross-domain recommendation lies in effectively transferring knowledge and patterns across domains while accounting for variations in user behaviors and item characteristics. Techniques such as domain adaptation, transfer learning, and hybrid models are often employed to bridge the gaps between different domains and optimize recommendation performance.  Moreover, recent work in cross-domain offline RL would be beneficial.\citet{liu2023beyond} present BOSA (Beyond OOD State Actions), a method for cross-domain offline RL (RL). BOSA tackles the challenges of out-of-distribution (OOD) state actions and data inefficiency by incorporating additional source-domain data. The authors propose specific objectives to address OOD transition dynamics and demonstrate that BOSA improves data efficiency and outperforms existing methods. The method is also applicable to model-based RL and data augmentation techniques. However, in offline RL4RS, this problem is still open for investigation as the techniques mentioned have not yet been explored in offline RL4RS.

\subsection{Implicit Feedback and Large Language Models}
Implicit feedback serves as a commonly employed feedback mechanism for learning recommendation policies in RS. Implicit feedback encompasses user actions like clicks, views, purchases, time spent, and dwell time during interactions with platforms or systems, signifying user preferences and interests. Although not as explicit as ratings or reviews, these behaviors offer valuable insights. In the context of RL4RS, the reward mechanism evaluates recommended items. Typically, this involves binary rewards based on click behavior, with some efforts, like \citet{zheng2018drn}, incorporating dwell times for a more comprehensive reward signal. However, accommodating multiple implicit feedback sources concurrently in RL4RS poses challenges due to limited relevant datasets or simulations. Additionally, harnessing review comments, a common type of implicit feedback in RS, within RL4RS remains a subject of exploration. \citet{zhang2019text} propose a text encoder solution, albeit relying on a manually gathered generator to produce review texts, which primarily validate feature learning rather than directly influencing the final reward. Transitioning this approach to offline RL4RS presents difficulties. Firstly, integrating review comments into the reward function requires careful study. Secondly, textual data introduces high-dimensional state representations, potentially necessitating novel algorithms tailored to this scenario.

Recently, Large Language Models (LLMs) have received increasing research interest in RS. LLM demonstrates a superior capability in handling textual data from multiple tasks such as natural language understanding, contextual understanding and sentiment analysis~\cite{zhao2023survey}. Existing RS works provides some insights about how LLMs can be adopted in RS such as prompt engineering to instruct the LLM to make recommendations~\cite{zhang2023recommendation}, utilizing the Generative Pre-trained Transformer (GPT) as the backbone to process features~\cite{sun2019bert4rec} etc. 

Moreover, some attempts have been undertaken about how LLM can be used in RL.\citet{du2023guiding} introduce a method called ELLM (Exploring with Large Language Models) that aims to enhance pretraining in RL by using LLM. ELLM works by prompting an LLM with a description of the agent's current state and then rewarding the agent for achieving goals suggested by the LLM. This method biases exploration towards behaviors that are meaningful and potentially useful from a human perspective, without needing human intervention. Meanwhile,\citet{carta2023grounding} explore the use of LLM in interactive environments through an approach called GLAM (Grounding Language Models). This method aligns the knowledge of LLMs with the environment, focusing on aspects like sample efficiency, generalization to new tasks, and the impact of online RL.

\subsection{Causality}
In the previous section, we mentioned that offline RL can be formulated as answering counterfactual queries. It is an intuitive choice to integrate the causality into  offline RL from this perspective. Moreover, causality is widely used in RS and receiving increasing interest in offline RL. We believe it would be a promising topic in offline RL4RS.

In the work by \citet{zhu2022offline}, an exploration is undertaken regarding the integration of causal world models into the domain of model-based offline RL. The theoretical underpinning of their study accentuates the superiority of causal world models over ordinary world models in the context of offline RL. This advantage is attributed to the incorporation of causal structure within the generalization error bound. The authors introduce an operational algorithm termed FOCUS (Offline Model-based RL with Causal Structure) to exemplify the potential value derived from comprehending and effectively utilizing causal structure in the domain of offline RL.

Additionally, \citet{liao2021instrumental} introduce the notion of instrumental variable value iteration for causal offline RL. The presentation of their work introduces IV-aided Value Iteration (IVVI), an algorithm designed with efficiency in mind, aimed at extracting optimal policies from observational data in the presence of unobserved variables. The utilization of instrumental variables (IVs) forms the foundation, with the authors devising a framework named Confounded Markov Decision Process with Instrumental Variables (CMDP-IV) to contextualize the problem. Notably, the IVVI algorithm, established upon a primal-dual reformulation of a conditional moment restriction, emerges as the first demonstrably efficient solution for instrument-aided offline RL.

One of the most common applications of integrating causality into the RL4RS is counterfactual augmentation.\citet{chen2022empowerment,chen2023intrinsically} develop a data augmentation technique that employs counterfactual reasoning to produce more informative interaction trajectories for RL4RS.\citet{wang2022causal} introduces the Causal Decision Transformer for RS (CDT4Rec), a model that merges offline RL with the transformer architecture. CDT4Rec is designed to tackle the challenges of crafting reward functions and leveraging large-scale datasets in RS. It employs a causal mechanism to deduce rewards from user behavior and uses the transformer architecture to handle vast datasets and identify dependencies.

Drawing inspiration from the works mentioned above, exploring causality in offline RL4RS emerges as a promising avenue for future research. Particularly, as causal offline RL4RS advances, its primary emphasis on counterfactual augmentation highlights an exciting direction. However, it is important to recognize the need for additional endeavors in different domains, including but not limited to distribution shifts and the presence of biases.

\subsection{Robustness}
The vulnerability of deep learning-based methods is evident through adversarial samples, underscoring the pressing concern of robustness in both RS and RL. Particularly, the exploration of adversarial attacks and defense strategies within the domain of RS has garnered significant attention in recent times, as emphasized by the comprehensive survey conducted by~\cite{deldjoo2021survey}. This attention is fueled by the critical importance of security within the realm of RS operations.

Furthermore, the vulnerability of RL policies to adversarial perturbations in agents' observations has been established by~\cite{lin2017tactics}. In the context of RL4RS,\citet{cao2020adversarial} introduce an adversarial attack detection approach. This method leverages the utilization of a Gated Recurrent Unit (GRU) to encode the action space into a lower-dimensional representation, alongside the design of decoders to identify potential attacks. However, it's important to note that this method exclusively addresses attacks rooted in the Fast Gradient Sign Method (FGSM) and strategically-timed maneuvers. As a result, its ability to detect other forms of attacks is limited.

Within the arena of offline RL, recent advancements provide a promising direction.\citet{panaganti2022robust} address the challenge of robust offline RL, centering on the learning of policies that can withstand uncertainties in model parameters. The authors introduce the Robust Fitted Q-Iteration (RFQI) algorithm, which relies solely on offline data to determine the optimal robust policy. This algorithm adeptly tackles concerns such as offline data collection, model optimization, and unbiased estimation. Additionally,\citet{zhang2022corruption} concentrate on a scenario involving a batch dataset of state-action-reward-next state tuples, susceptible to potential corruption by adversaries. Their objective is to extract a near-optimal policy from this compromised dataset.

\section{Conclusion}
The recent advancements in RL4RS pave the way for efficiently capturing users' dynamic interests. However, the nature of online interactions necessitates costly trajectory collection procedures, posing a significant hurdle for researchers interested in this field.
In this survey, our goal is to provide a comprehensive overview of offline RL4RS, a novel paradigm that eliminates the need for an expensive data collection process. Alongside reviewing recent works, we also offer insights into potential future opportunities. Specifically, we've compiled and analyzed recent progress in offline RL4RS, organized into two distinct categories: off-policy learning utilizing logged data and offline RL4RS techniques.
Furthermore, we address several prevailing challenges in this domain: offline data quality, distribution shift, bias and variance, and explainability. Additionally, we present potential avenues for future exploration in this rapidly evolving field, such as cross-domain recommendation, LLMs, causality, and robustness.
Being an emerging topic, offline RL4RS introduces fresh possibilities for integrating pre-existing offline datasets into the realm of RL4RS. This survey can also be perceived as a visionary paper, offering potential benefits to researchers who are newcomers to this field.

\bibliographystyle{ACM-Reference-Format}
\bibliography{sample-base}

\end{document}